\documentclass[final,1p,times]{elsarticle}
\usepackage{txfonts}
\usepackage{graphicx}
\usepackage{natbib}

\usepackage{amssymb}

\journal{Physica A}

\begin{document}

\begin{frontmatter}



\title{Quark deconfinement in protoneutron star cores:  effect of color superconductivity within the MIT bag model}
\author{T. A. S. do Carmo  and G. Lugones}

\address{Universidade Federal do ABC, Rua Santa Ad\'elia, 166, 09210-170, Santo Andr\'e, Brazil}



\begin{abstract}
We analyze the effect of color superconductivity in the transition from hot hadron matter to quark matter in the presence of a gas of trapped electron neutrinos. To describe  strongly interacting matter we adopt a two-phase picture in which the hadronic phase is described by means of a  non-linear Walecka model and just deconfined matter through the MIT bag model including color superconductivity. We impose flavor conservation during the transition in such a way that just deconfined quark matter is transitorily out of equilibrium with respect to weak interactions. Our results show that color superconductivity facilitates the transition for temperatures below $T_c$. This effect may be strong if the superconducting gap is large enough. As in previous work we find that trapped neutrinos increase the critical density  for deconfinement;  however, if the just deconfined phase is color superconducting this effect is weaker than if deconfined matter is unpaired. We also explore the effect of different parametrizations of the hadronic equation of state (GM1 and NL3) and the effect of hyperons in the hadronic phase. We compare our results with those previously obtained employing the Nambu-Jona-Lasinio model in the description of just deconfined matter and show that they are in excellent agreement if the bag constant $B$ is properly chosen.
\end{abstract}

\begin{keyword}
phase transition \sep quark matter \sep color superconductivity \sep protoneutron stars
\end{keyword}

\end{frontmatter}


\section{Introduction}

Among the most important unsolved questions concerning the behavior of matter inside neutron
stars, it is the knowledge of the thermodynamic conditions at which the deconfinement transition to quark matter would occur  \cite{Logoteta,Hempel,Lugones2009,LugonesCarmoNJL,Fraga2,Fraga}. The analysis of such problem is complicated by the uncertainties in the knowledge of the equation of state (EoS) above the nuclear saturation density, as well as by the lack of a satisfactory description of both hadronic a deconfined matter within a unified description (however see \cite{Lawley2006,Lawley2006a,Dexheimer} for work in this direction).
A possible approach, is to analyze the transition within a two-phase description in which an hadronic model valid around the nuclear saturation density $\rho_{0}$ is extrapolated to larger densities and a quark model that is expected to be valid only for asymptotically large densities is extrapolated downwards. Within this kind of analysis some work has been performed recently in order to determine the effect of different hadronic and quark equations of state, as well as the effect of different astrophysical environments: e.g. different temperatures, effect of color superconductivity, effect of neutrino trapping \citep{Lugones2005,Logoteta,Lugones2009,LugonesCarmoNJL}.

An important characteristic of the deconfinement transition in neutron stars, is that quark and lepton flavors must be conserved during the transition  \citep{Madsen1994,IidaSato1998,Lugones1998,Lugones1999,Bombaci2004,Lugones2005,Bombaci2007,Logoteta,Lugones2009,LugonesCarmoNJL}.
As a consequence, just deconfined quark matter is transiently out of equilibrium with respect to weak interactions (for a short period of $\sim 10^{-7}$ s).
When color superconductivity is included together with flavor conservation, the most likely configuration of the just deconfined
phase is two-flavor color superconducting (2SC) provided the pairing gap is large enough \citep{Lugones2005}.
In a recent paper \citep{LugonesCarmoNJL} we have investigated the role of color superconductivity in the deconfinement transition in protoneutron star (PNS) conditions employing the Nambu-Jona-Lasinio (NJL) model for just deconfined quark matter.
While early calculations showed that neutrino trapping may preclude the transition in PNSs  \citep{Lugones1998,Lugones1999}, the results in Ref. \citep{LugonesCarmoNJL}  show that color superconductivity compensates this effect resulting a transition density that is nearly constant throughout the deleptonization stage.  However, a full calculation including both effects has not been done yet within the MIT bag model.

In the present paper we shall analyze the deconfinement transition
in protoneutron star conditions employing for hadronic matter a nonlinear Walecka model which includes hadrons,  electrons,  and  electron  neutrinos  in  equilibrium  under weak interactions. For the just deconfined quark matter
we shall employ the MIT bag model including the effect of color superconductivity and neutrino trapping.
According to numerical simulations \citep{Pons}, during the first tens of seconds of evolution the protoneutron star cools from $T \sim 40-50$ MeV to temperatures below $2$ MeV. In the same period, the chemical potential of the trapped neutrinos evolves from $\sim 150-200$ MeV to essentially zero.
This paper extends previous calculations performed in Ref. \citep{Taiza_JPG2013}. In that paper we showed that the transition density for bubbles with radii around $\sim 100 - 200$ fm is almost coincident with the bulk transition density and that the nucleation rate of such droplets is huge. This fully justifies a more detailed treatment of several aspects of the deconfinement transition, without including surface effects at all.
In view of this, we investigate here the effect of other parametrizations of the hadronic equation of state (GM1 and NL3), the effect of hyperons in the hadronic phase, we explore more parameters of the quark model, and analyze in more detail the effect of trapped neutrinos and color superconductivity.

The article is organized as follows. In Sec.~2 we present the main aspects of the equations of state. In Sec.~3 we study
the deconfinement transition at finite temperature for different parametrizations of the equations of state and different neutrino
trapping conditions. In Sec. 4 we discuss our results and compare them with previous calculations using the NJL model in the description of quark matter.

%
\section{Equations of state} \label{EOS}
\subsection{Hadronic matter}\label{HM}
For the hadronic phase we use a non-linear Walecka model \citep{Walecka,Walecka2,Boguta1977,Glendenning1991}
for matter with and without hyperons. For matter with hyperons  we include the whole baryon octet, electrons, electron neutrinos, and the corresponding antiparticles. For matter  with no hyperons we consider nucleons, electrons, electron neutrinos, and the corresponding antiparticles. The Lagrangian is given by
\begin{equation}
{\cal L}={\cal L}_{B}+{\cal L}_{M}+{\cal L}_{L}, \label{octetlag}
\end{equation}
where the indices $B$, $M$ and $L$ refer to baryons, mesons and
leptons respectively. For the baryons we have
\begin{eqnarray}
{\cal L}_B= \sum_B \bar \psi_B \bigg[\gamma^\mu\left
(i\partial_\mu - g_{\omega B} \ \omega_\mu- g_{\rho B} \ \vec \tau
\cdot \vec \rho_\mu \right)
-(m_B-g_{\sigma B} \ \sigma)\bigg]\psi_B,
\end{eqnarray}
with $B$ extending over the nucleons $N= n$, $p$ and the following
hyperons $H = \Lambda$, $\Sigma^{+}$, $\Sigma^{0}$, $\Sigma^{-}$,
$\Xi^{-}$, and $\Xi^{0}$. The contribution of the mesons $\sigma$,
$\omega$ and $\rho$ is given by
\begin{eqnarray}
{\cal L}_{M} &=& \frac{1}{2} (\partial_{\mu} \sigma \ \!
\partial^{\mu}\sigma -m_\sigma^2 \ \! \sigma^2) - \frac{b}{3} \ \!
m_N\ \! (g_\sigma\sigma)^3 -\frac{c}{4} \ (g_\sigma \sigma)^4
\nonumber\\
& & -\frac{1}{4}\ \omega_{\mu\nu}\ \omega^{\mu\nu} +\frac{1}{2}\
m_\omega^2 \ \omega_{\mu}\ \omega^{\mu}
-\frac{1}{4}\ \vec \rho_{\mu\nu} \cdot \vec \rho\ \! ^{\mu\nu}+
\frac{1}{2}\ m_\rho^2\  \vec \rho_\mu \cdot \vec \rho\ \! ^\mu,
\end{eqnarray}
where the coupling constants are
\begin{eqnarray}
g_{\sigma B}=x_{\sigma B}~ g_\sigma,~~g_{\omega B}=x_{\omega B}~
g_{\omega},~~g_{\rho B}=x_{\rho B}~ g_{\rho}.
\end{eqnarray}
Electrons and neutrinos are included as a free Fermi gas, ${\cal L}_{L}=\sum_l \bar \psi_l \left(i \rlap/\partial -
m_l\right)\psi_l$, in chemical equilibrium with all other particles.
For details on the explicit form of the equation of state derived from this Lagrangian the reader
is referred to Ref. \citep{Taiza_JPG2013}.
The equation of state  can be solved numerically by  specifying
three thermodynamic quantities, e.g. the temperature $T$, the mass-energy density $\rho$
and the chemical potential of electron neutrinos in the hadronic phase $\mu_{\nu_e}^H$.
The constants in the model are determined by the
properties of nuclear matter and hyperon potential depths
known from hypernuclear experiments \citep{Friedman,Weissenborn}.
In the present work we use the GM1 parametrization given by \cite{Glendenning1991}
and the NL3 parametrization given by \cite{Lalazissis}, as shown in Table~\ref{table1}.
For each parametrization we construct an equation of state including nucleons plus leptons and another one including the baryon octet plus leptons. They are labeled as GM1npe$+\nu$, GM1hyp$+\nu$, NL3npe$+\nu$ and NL3hyp$+\nu$. The  maximum masses $M_{max}$ of hadronic stars are 1.78 $M_{\odot}$ for GM1npe$+\nu$, 2.32 $M_{\odot}$ for GM1hyp$+\nu$, 1.95 $M_{\odot}$  for NL3npe$+\nu$ and 2.7 $M_{\odot}$ for NL3hyp$+\nu$. Except for GM1npe$+\nu$, these masses are compatible with the masses of the pulsars PSR J1614-2230 with $1.97 \pm 0.04 M_\odot$ \citep{Demorest} and PSR J0348+0432 with $2.01 \pm 0.04 M_\odot$ \cite{Antoniadis2013}.
The parametrization for the hyperon coupling constants in the case of GM1 is $x_{\omega i} =  0.666$,
$x_{\sigma i} =  0.6104 $ and  $x_{\rho i}  = 0.6104$ \citep{Glendenning1991}. For NL3 we use $x_{\omega \Lambda}  = x_{\omega \Sigma} = 0.6666$,  $x_{\omega \Xi}  = 0.3333$, $x_{\sigma \Lambda}  = 0.6106 $,  $x_{\sigma \Sigma}  = 0.4046$,  $x_{\sigma \Xi}   = 0.3195$ and   $x_{\rho i}  = 1$  \citep{Chiapparini2009}.

\begin{table}[t]
\centering
\begin{tabular}{c|c|c|c|c|c}
\hline\hline
\quad Label \quad & \quad $\left({g_{\sigma}}/{m_{\sigma}}\right)^{2}$ \quad & \quad $\left({g_{\omega}}/{m_{\omega}}\right)^{2}$ \quad & \quad  $\left({g_{\rho}}/{m_{\rho}}\right)^{2}$ \quad & \quad b \quad & \quad c  \quad   \\
  \hline
\quad   GM1 \quad  & 11.79 fm$^{2}$ & 7.149 fm$^{2}$ & 4.411 fm$^{2}$ & 0.002947 & -0.001070 \\
\quad   NL3  \quad & 15.8 fm$^{2}$ & 10.51 fm$^{2}$ & 5.35 fm$^{2}$ & 0.002052 & -0.002651  \\
\hline\hline
\end{tabular}
\caption{Parameters of the hadronic equation of state.  The parametrizations for the hyperon coupling constants
and the maximum masses of hadronic stars are given in the text. }
\label{table1}
\end{table}

\subsection{Quark matter}\label{QM}

The quark phase is composed by \textit{u}, \textit{d}, and \textit{s} quarks, electrons, electron neutrinos and the corresponding antiparticles.
We describe this phase by means of the MIT bag model at finite temperature with zero strong coupling constant, zero \textit{u} and \textit{d} quark masses and strange quark mass $m_{s} = 150$ MeV.

The total thermodynamic potential can be written as:
\begin{equation}
\Omega = \Omega_{Q}  +  \Omega_{L} +  B ,\label{omega}
\end{equation}
where the indexes $Q$ and $L$ refer respectively to quarks and leptons. The contribution of quarks is given by
$\Omega_{Q} = \sum \Omega_{cf}$,  being $f  = u, d, s$   the flavor index and $c = r, g, b$  the color index.
For free  unpaired quarks we employ
\begin{eqnarray}
\Omega_{cf} = - \frac{\gamma T}{2 \pi^{2}}\int^{\infty}_{0}k^{2}\ln\left[1 + e^{-\left(\frac{E_{f}-\mu_{cf}}{T}\right)}\right]dk, \label{omega_q}
\end{eqnarray}
being $E_{f} = \sqrt{k^{2}+ m^{2}_{f}}$ the particle energy, and $\mu_{cf}$ the particle chemical potential (we use $-\mu_{cf}$ for antiparticles).
For paired quarks we use the expression:
\begin{eqnarray}
\Omega_{cf} = - \frac{\gamma T}{2 \pi^{2}}\int^{\infty}_{0}k^{2}\ln\left[1 + e^{-\frac{\varepsilon_{cf}}{T}}\right]dk, \label{omega_delta}
\end{eqnarray}
where $\varepsilon_{cf} = \pm \sqrt{(E_{f}-\mu_{cf})^{2}+ \Delta^{2}}$ is the single-particle energy dispersion relation  when it acquires an energy gap $\Delta$.
Notice that, for particles, we can obtain Eq.~(\ref{omega_q}) from Eq.~(\ref{omega_delta}) in the limit $\Delta = 0$ by considering the minus sign in the dispersion relation for $E_{f} < \mu_{cf}$ and the plus sign $E_{f} > \mu_{cf}$ (see e.g. \cite{Schmitt2010}).

The gap equations for a color-superconducting condensate
with total spin $J=0$ have been derived  perturbatively in dense QCD \citep{Pisarski1999}.
To leading order in the weak coupling, the temperature dependence of the condensate is
identical to that in BCS-like theories \citep{Pisarski1999}; thus,
we employ the following temperature dependence of the gap parameter of Eq.~(\ref{omega_delta})
\begin{eqnarray}
\Delta(T) = \Delta_{0}\sqrt{1- \left(\frac{T}{T_{c}}\right)^{2},} \label{delta(T)}
\end{eqnarray}
where the critical temperature for the 2SC phase is $T_{c} = 0.57 \Delta_{0}$ \citep{Son1999,SchmittWang}.

The contribution of leptons is given by $\Omega_{L} = \sum_{i} - \frac{\gamma T}{2 \pi^{2}}\int^{\infty}_{0} dk \, k^{2}\ln [1 + \exp (-[E_i -\mu_i] /T)]$ with $i = e^-, e^+, \nu_e, \bar{\nu}_e $ and   $E_i = (k^2+ m_i^2)^{1/2}$.
The degeneracy factor is   $\gamma = 2, 2, 1$ for quarks, electrons and  neutrinos respectively. In all cases, the contribution of antiparticles is obtained through $\bar{\mu} = - \mu$.
From the grand thermodynamic potential we obtain the pressure $P$, the quark number density of each flavor and color $n_{fc}$, the number density of electrons $n_{e}$, and of electron neutrinos $n_{\nu_{e}}$. The Gibbs free energy per baryon is
$g = ( \sum_{fc} \mu_{fc} n_{fc} + \mu_{e} n_{e} + \mu_{\nu_{e}} n_{\nu_{e}} ) / n_B$.
As we shall see in the next section, the here-considered \textit{just deconfined} phase is out of chemical
equilibrium with respect to weak interactions. Thus, the chemical potentials $\mu_{cf}$, $\mu_{e}$ and $\mu_{\nu_{e}}$ are obtained from the hadronic phase through flavor conservation conditions.

%
\begin{figure}[t]
\centering
\includegraphics[scale=0.42]{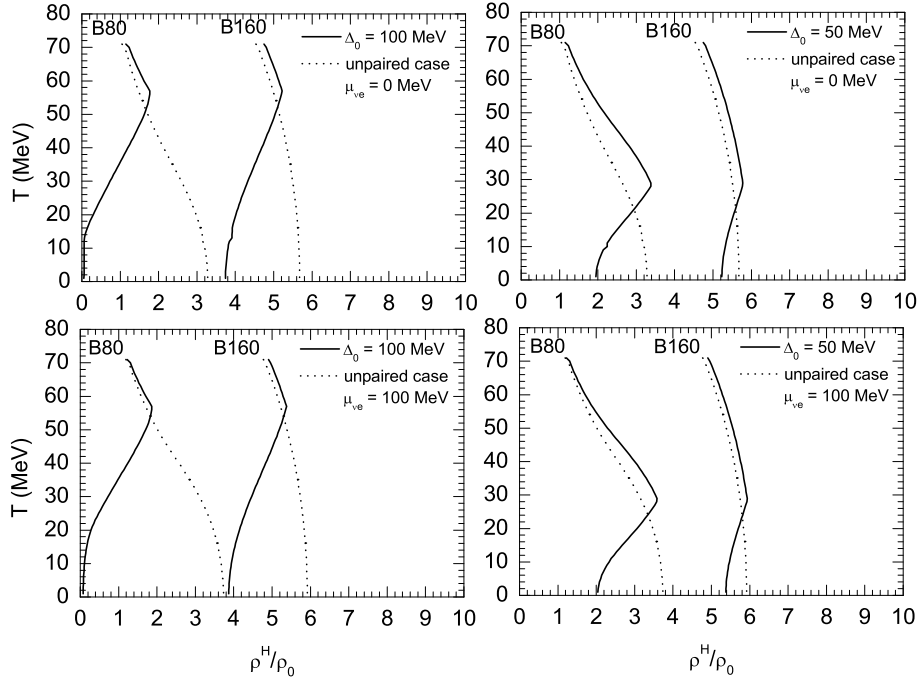}
\caption{The hadron-matter mass-energy density at which the deconfinement phase transition occurs as a function of the temperature T. Density is given in units
of the nuclear saturation density $\rho_{0}$, being $\rho_{0} = 2.7 \times 10^{14}$ g/cm$^{3}$. We employed the GM1npe$+\nu$ hadronic EoS. The results are shown for quark matter without pairing (dotted lines) and for color superconducting quark matter with $\Delta_{0} = $ $50$ and $100$ MeV (full lines). The adopted values for the bag constant are B $= 80$ MeV/fm$^{3}$ and  B $= 160$ MeV/fm$^{3}$ and are
indicated with the labels B80 and B160 respectively. We employed different values for the chemical potential of the trapped neutrinos in the hadronic phase: $\mu_{\nu e} =0, 100$ MeV. Notice that in general electron neutrinos push the transition density upwards, but in the case of color superconducting matter the effect is very slight. The decrease of the transition density due to color superconductivity is clearly seen in solid lines for  $T < T_c$. }
\label{1}
\end{figure}

Concerning the bag constant $B$,  we shall employ $B=80$ MeV/fm$^{3}$ corresponding to absolutely stable quark matter (strange matter) and  $B=160$ MeV /fm$^{3}$ corresponding to quark matter that is allowed only at high pressures. We shall also employ very large values of $B$ in order to establish a comparison with previous work performed within the frame of the NJL model \citep{LugonesCarmoNJL}. As discussed in \cite{buballa2005} it is possible to define within the NJL model a quantity $B'$ which is the difference between the pressure of the interacting quark matter and that of the free one, both taken at vanishing temperature and chemical potential. Thus, $B'$ plays a role similar to that of the bag constant $B$ in the MIT bag model. However, notice that in the NJL model $B'$ is \textit{not} an extra parameter since its value can be determined once the model parameters are fixed. In this work we shall employ the values $B$ = 353 MeV/fm$^{3}$  and 337.2 MeV/fm$^{3}$ because they correspond to the parametrizations  \textit{set 1} and \textit{set 2} of the NJL model employed in \cite{LugonesCarmoNJL}.

%
\begin{figure}[t]
\begin{center}
\includegraphics[scale=0.42]{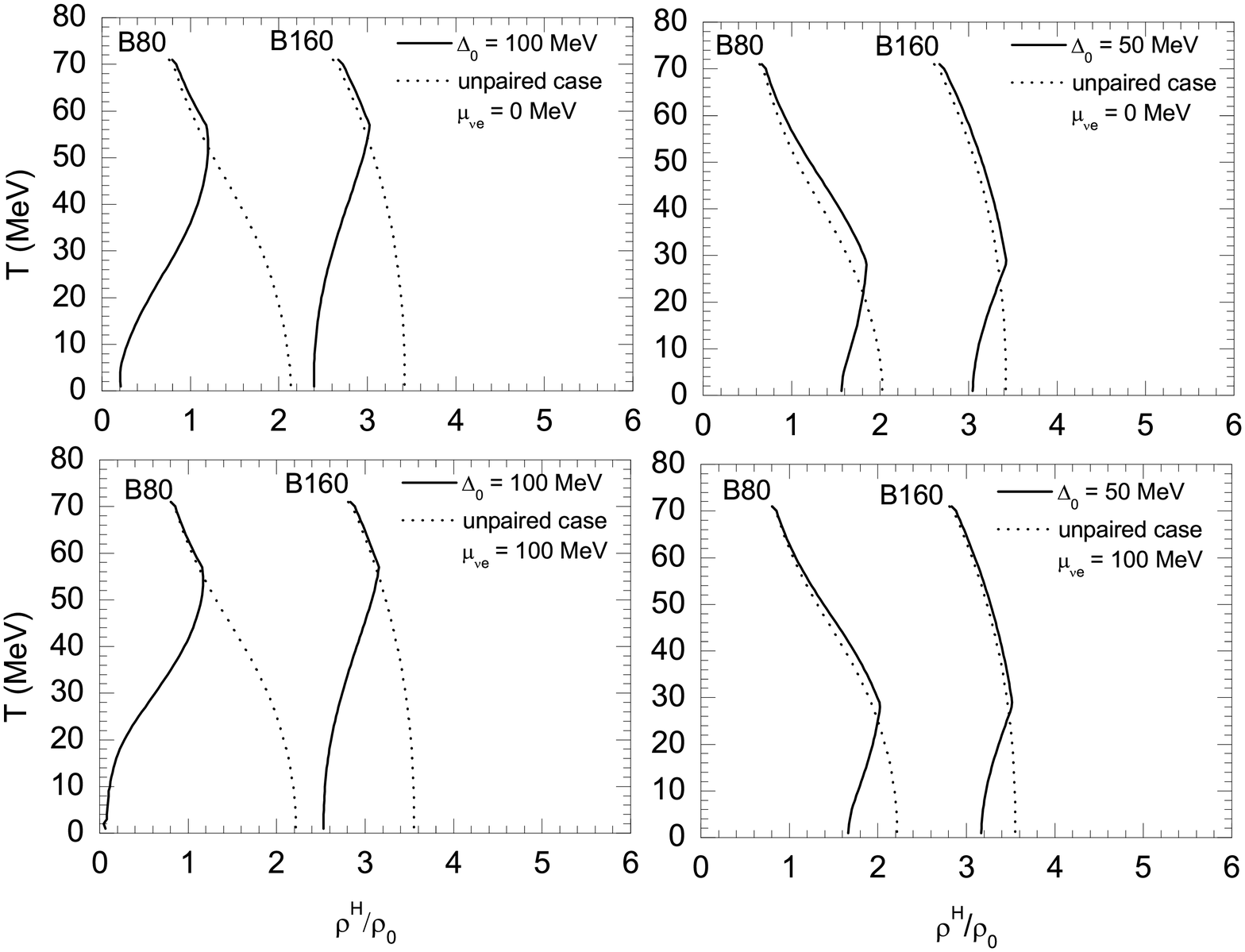}
\end{center}
\caption{Same as Fig.~1 but employing the NL3hyp$+\nu$ hadronic EoS.
Notice that the color-superconducting phase does not reduce to the normal quark phase above the critical temperature (gap equal to zero). This happens because the gap $\Delta_0$ is fixed to a constant and the pairing pattern is obtained by construction, in a similar way as in \cite{pattern,pattern2,Lugones2002,pattern4,pattern5}. While a more refined model may allow to find the curve that smoothly connects the full and the dashed lines in Figs.~1 and 2 for temperatures around the critical one, none of the main conclusions of the present manuscript are expected to change with such calculation.
The mass-energy density  \textit{in the quark phase} just after deconfinement can also be obtained from our calculations, and can be represented as in Figs.~1 and 2. However, since this density changes in roughly $10^{-7} s$ due to weak interactions we  indicate just the hadron density where the transition begins. The determination of the final density after weak equilibration of quark matter involves the solution of a Boltzmann equation together with the implementation of the rates of weak processes among all the particles. This is out of the scope of the present work.}
\label{2}
\end{figure}

\section{Deconfinement transition}

To calculate the conditions for the transition standard Gibbs' criterion is applied, i.e. equality of pressure \textit{P}, temperature \textit{T} and Gibbs free energy per baryon \textit{g} in both the hadronic (H) and the quark (Q) phases \footnote{In any system that is in contact with a temperature and pressure reservoir, a thermodynamic process can happen spontaneously if the Gibbs free energy of the final state is smaller or equal than the Gibbs free energy of the initial state. This is derived from the second law of thermodynamics for an isolated system.
In our case, the final state is in equilibrium with respect to strong interactions, i.e. it is an equilibrium state for timescales much shorter than a typical scale of weak interactions (i.e. $\ll 10^{-7}$ s). Since the system has a well-defined free energy within these timescales, the Gibbs version of the second law of thermodynamics can be applied.} :
\begin{eqnarray}
T^{H} &=& T^{Q}, \;\;\; \\
P^{H}(T^{H}, \mu_{p}, \mu_{\nu_{e}}^{H}) &=& P^{Q}(T^{Q}, \{ \mu_{fc} \},  \mu_{e}^{Q}, \mu_{\nu_{e}}^{Q}), \label{dec01} \;\;\; \\
g^{H}(T^{H}, \mu_{p}, \mu_{\nu_{e}}^{H})&=& g^{Q}(T^{Q}, \{ \mu_{fc} \}, \mu_{e}^{Q}, \mu_{\nu_{e}}^{Q}). \label{dec02} \;\;\;
\end{eqnarray}
On the other hand, deconfinement is driven by strong interactions and therefore quark and lepton flavors must be conserved during the
deconfinement transition \citep{Madsen1994,IidaSato1998,Lugones1998,Lugones1999,Bombaci2004,Lugones2005,Bombaci2007}.
When a small quark-matter drop is nucleated at the core of a compact star, the abundances of all particle species inside
it must be \textit{initially} the same as
in the hadronic matter from which it has been originated. Thus we have
\begin{eqnarray}
Y_{i}^{H}(T^{H}, \mu_{p}, \mu_{\nu_{e}}^{H}) &=& Y_{i}^{Q}(T^{Q}, \{ \mu_{fc} \}, \mu_{e}^{Q}, \mu_{\nu_{e}}^{Q}),  \label{dec03}
\end{eqnarray}
with $i = u, d, s, e, \nu_e$, being  $Y^H_i \equiv n^H_i / n^H_B$ and  $Y^Q_i \equiv n^Q_i /
n^Q_B$ the abundances of each particle species in the hadronic and quark
phase respectively.
The number densities of $u, d$ and  $s$ quarks in the hadronic phase are given by
$n^H_u = 2 n_p  +   n_n  + n_{\Lambda} + n_{\Sigma^{0}} +  2
n_{\Sigma^{+}}  + n_{\Xi^{0}}$,  $n^H_d =  n_p  +  2  n_n  + n_{\Lambda} + n_{\Sigma^{0}} +  2
n_{\Sigma^{-}}  + n_{\Xi^{-}} $ and $n^H_s =n_{\Lambda}  + n_{\Sigma^{+}} + n_{\Sigma^{0}}  +
n_{\Sigma^{-}} + 2 n_{\Xi^{0}} + 2 n_{\Xi^{-}}$.

Notice that, since the hadronic phase is assumed to be
electrically neutral, flavor conservation ensures automatically
the charge neutrality of the just deconfined quark phase.
We also emphasize that the here-studied \textit{just deconfined} drop is transitorily out of equilibrium
under weak interactions. After some time (typically $\tau_{weak} \sim 10^{-7}$ s) weak interactions
drive this drop to a $\beta$-stable configuration. However, in spite of being very short lived, the here
studied phase determines whether the hadronic phase may deconfine or not.

Additionally, the deconfined phase must be locally colorless, i.e. it must be composed by an equal
number or \textit{red}, \textit{green} and \textit{blue} quarks:
\begin{eqnarray}
n_{r}(T^{Q}, \{ \mu_{fr} \}) &=& n_{g}(T^{Q}, \{ \mu_{fg} \}) , \label{dec04}\\
n_{r}(T^{Q}, \{ \mu_{fr} \}) &=& n_{b}(T^{Q}, \{ \mu_{fb} \}) . \label{dec05}
\end{eqnarray}

Finally, it has been shown that when color superconductivity is included together with flavor
conservation and color neutrality, the most likely configuration of the just deconfined
phase is 2SC provided the pairing gap is large enough \citep{Lugones2005}.
Thus, in order to allow for pairing between quarks $d_{r}$ with $u_{g}$ and between quarks $u_{r}$ with $d_{g}$
we impose that\footnote{We are assuming a conventional 2SC phase for which the chemical potential of the quarks that form a Cooper pair are equal \citep{AlfordSchmitt}. Since both quarks are considered to be massless this implies the equality of the corresponding number densities. The formation of gapless 2SC phases with a mismatch between the Fermi surfaces of the pairing $u$ and $d$ quarks \citep{HuangShovkovy}, also deserve study. However, they are not considered here because within the present treatment this would require the introduction of an additional free parameter (i.e. the difference between the Fermi momenta of $u$ and $d$ quarks). }:
\begin{eqnarray}
n_{ur}(T^{Q}, \mu_{ur}) &=& n_{dg}(T^{Q}, \mu_{dg}) , \label{dec06} \\
n_{dr}(T^{Q}, \mu_{dr}) &=& n_{ug}(T^{Q}, \mu_{ug}). \label{dec07}
\end{eqnarray}

Equations (\ref{dec04}-\ref{dec07}) together with the assumption that $\mu_{sr} = \mu_{sb}$
force a 2SC pattern for the just deconfined phase (see \cite{Lugones2005} for more details). Notice that for a given value of the temperature and
the chemical potential of the neutrinos in the hadronic phase $\mu_{\nu_{e}}^{H}$, equations (\ref{dec01}-\ref{dec07})
univocally determine the pressure at which deconfinement occurs. Also notice that, according to the present description,
the pressure $P$ and the Gibbs free energy per baryon $g$ are the same in both the hadronic phase and the just deconfined phase.
However, the mass-energy densities $\rho^{H}$ and $\rho^{Q}$ are different in general.
Similarly, while the abundance $Y_{\nu_e}$ of neutrinos is the same in both the hadronic and just  deconfined quark phases,
the chemical potentials $\mu_{\nu_e}^Q$ and $\mu_{\nu_e}^H$ are different.

%
\begin{figure}[t]
\centering
\includegraphics[scale=0.42]{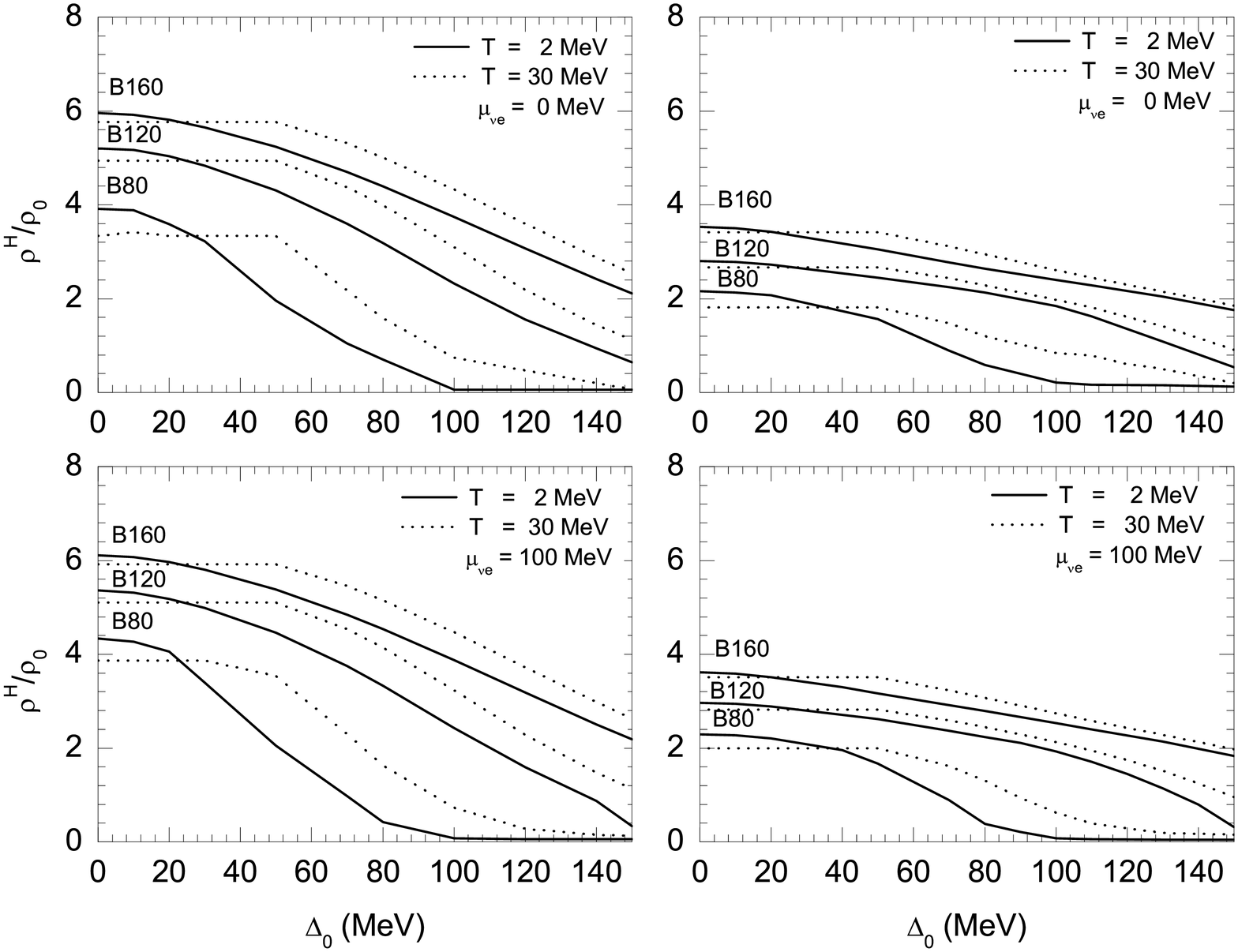}
\caption{Mass-energy density of hadronic matter at which deconfinement occurs as a function of $\Delta_{0}$ for the GM1npe$+\nu$ EoS (left panel) and the NL3hyp$+\nu$ EoS (right panel). We employed $\mu_{\nu_{e}} = 0$ MeV and  $\mu_{\nu_{e}} = 100$ MeV for both parametrizations. The values of the bag constant are B $= 80$ MeV/fm$^{3}$ (B80), B $= 120$ MeV/fm$^{3}$ (B120) and B $= 160$ MeV/fm$^{3}$ (B160). There is a strong decrease of the transition density $\rho^{H}/\rho_{0}$ for large enough $\Delta_{0}$.}
\label{3}
\end{figure}

In the following we analyze the effect on the deconfinement transition,  of temperature, neutrino trapping, color superconductivity, and hyperons in the hadronic EoS.

In Figs.~1 and 2 we show our results for the GM1npe$+\nu$ and NL3hyp$+\nu$ hadronic EoS respectively.
The two panels on the left correspond to the \textit{gap} values $\Delta_0 = 100$ MeV and the two panels on the right correspond to $\Delta_0 = 50$ MeV. The two upper panels of the figures correspond to the case without trapped neutrinos in hadronic matter ($\mu_{\nu_{e}}^{H} = 0$ MeV) while the two lower panels  correspond to hadronic protoneutron star matter with $\mu_{\nu_{e}}^{H} = 100$ MeV. In each panel of Figs.~1 and 2 we show four curves. The dotted curves correspond to quark matter without pairing and the full lines correspond to paired quark matter. Notice that the unpaired phase is not the same as the paired phase with $\Delta(T) = 0$ because the latter has a 2SC pattern while in unpaired quark matter all quarks with the same flavor have the same chemical potential independently of color. Since some energy must be paid in order to enforce a 2SC pattern, the full lines lie to the right of the dotted lines when $\Delta(T) = 0$ (i.e above $T_c$). However, for $T < T_c$ the curves for the paired case fall to the left of the curves for the unpaired case, i.e. color superconductivity shifts the transition density downwards. Since we have $T_c = 0.57 \Delta_{0}$ the effect begins to work at larger temperatures for larger $\Delta_{0}$.
Comparing the upper and lower panels of these figures we see that in most cases the curves with $\mu_{\nu_{e}}^{H} = 100$ MeV are shifted to the right with respect to the same curves but for $\mu_{\nu_{e}}^{H} = 0$ MeV. That is, in most cases trapped neutrinos push the transition density upwards as already shown by \cite{Lugones1998}, but in coincidence with \cite{LugonesCarmoNJL} the effect is very small when color superconductivity is taken into account.

The behavior of the transition density with the gap parameter $\Delta_0$ is shown in Fig.~3 for the GM1npe$+\nu$ and NL3hyp$+\nu$ hadronic EoS. The two upper panels of the figure correspond to $\mu_{\nu_{e}}^{H} = 0$ MeV and the two lower panels to $\mu_{\nu_{e}}^{H} = 100$ MeV.
The three full lines correspond to T = 2 MeV and the three dotted curves to T = 30 MeV. Each pair of curves correspond to three different values of the bag constant $B$.  Notice that the  mass-energy density of hadronic matter at which deconfinement occurs is a decreasing function of the gap parameter $\Delta_{0}$.
The effect is strong, e.g. the transition density for $\Delta_0$ = 100 - 150 MeV is much smaller than for $\Delta_0$ = 0 MeV.
For sufficiently small $\Delta_{0}$ the transition density $\rho^{H}$ has constant values. This is because this part of the curve corresponds to temperatures that are larger than the critical temperature $T_{c} = 0.57 \Delta_{0}$, and therefore the pairing gap $\Delta(T)$ is zero.

%
\begin{figure}[t]
\centering
\includegraphics[scale=0.42]{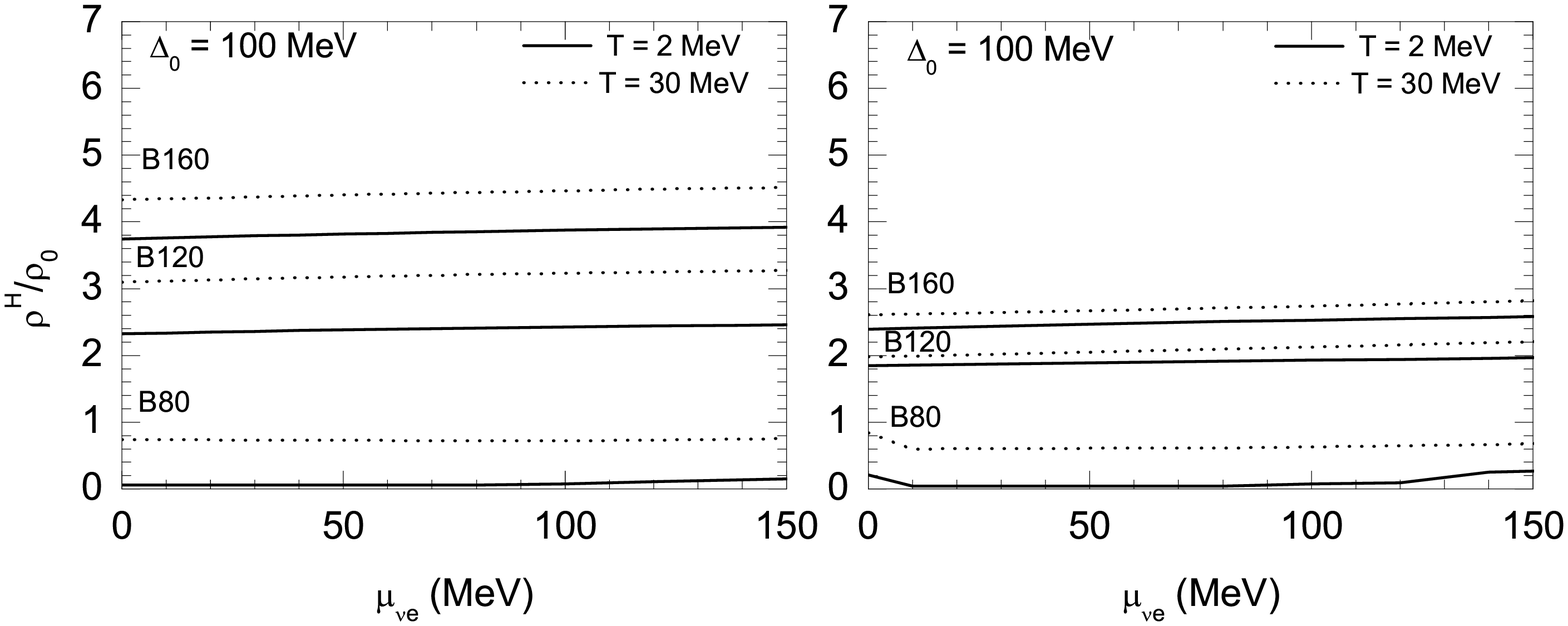}
\caption{Mass-energy density of hadronic matter at which deconfinement occurs versus the chemical potential of trapped electron neutrinos $\mu_{\nu_{e}}^{H}$ in the hadronic phase for the GM1npe$+\nu$ EoS (left panel) and the NL3hyp$+\nu$ EoS (right panel). }
\label{4}
\end{figure}
%
%
\begin{figure}[t]
\centering
\includegraphics[scale=0.42]{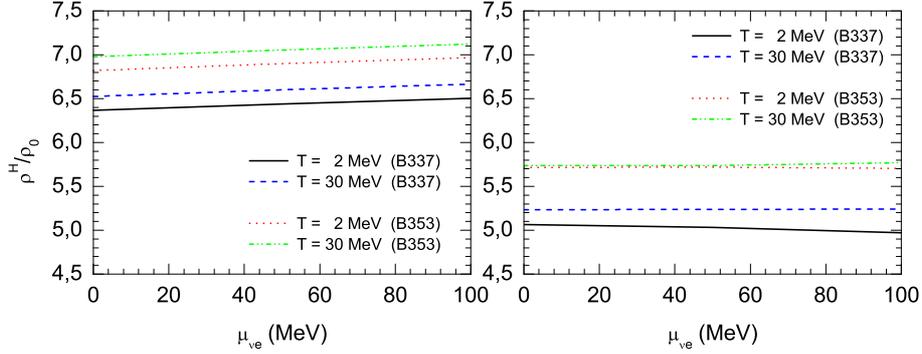}
\caption{Same as Fig.~4 but for  $B$ = 353 MeV/fm$^{3}$ (B353) and $B$ = 337 MeV/fm$^{3}$ (B337).
These values of $B$ correspond to the \textit{set 1} and \textit{set 2} parametrizations of the Nambu-Jona-Lasinio model respectively (see the discussion at the end of Sec.~2).}
\label{5}
\end{figure}

\begin{table}[h]
\centering
\begin{tabular}{l|c|c|c|c|c|c}
\hline \hline
 Bag  &   \multicolumn{2}{c}{$B$= 60 MeV fm$^{-3}$}  &\multicolumn{2}{c}{$B$= 120 MeV fm$^{-3}$}  &\multicolumn{2}{c}{$B$= 337 MeV fm$^{-3}$}   \\
\hline
  Temperature          & 2 MeV & 30 MeV  & 2 MeV & 30 MeV & 2 MeV  & 30 MeV\\
\hline \hline
GM1hyp$+\nu$   & 0.31  & 0.72 & 3.37 &  3.86 &  7.59  &  8.66  \\
GM1npe$+\nu$        & 0.55  & 0.74 & 4.06 &  4.74 &  6.76  &  7.28  \\
\hline
NL3hyp$+\nu$   & 0.24  & 0.64 & 2.59 &  2.80 &  5.04  &  5.66  \\
NL3npe$+\nu$        & 0.50  & 0.65 & 2.70 &  3.03 &  4.26  &  4.56  \\
\hline\hline
\end{tabular}
\caption{Effect of hyperons in the density for deconfinement. Density is given in units
of the nuclear saturation density $\rho_{0}$ for different values of the temperature, the bag constant and the two parametrizations of the hadronic EOS.  We have adopted $\mu_{\nu_{e}}^{H} = 100$ MeV and $\Delta_{0} = 100$ MeV.  }
\label{table_hyp}
\end{table}

The effect of the chemical potential of trapped neutrinos  $\mu_{\nu_{e}}^{H}$ is shown in Fig. 4  for $\Delta_{0}= 100$ MeV. Notice that the $\rho^H$ at which deconfinement occurs is a  slightly increasing function of $\mu_{\nu_{e}}^{H}$, that is, neutrino trapping tends to inhibit the transition but the effect is very small due to color superconductivity.

Fig.~5 is similar to Fig.~4 but for $B = 353$ MeV/fm$^{3}$  and $B = 337$ MeV/fm$^{3}$. These values of $B$ correspond respectively to the \textit{set 1} and \textit{set 2} parametrizations of the NJL model employed in \citep{LugonesCarmoNJL}. Notice that the deconfinement density  is very similar to the obtained in Fig.~3 of \cite{LugonesCarmoNJL}  within the frame of the NJL model.

In Table \ref{table_hyp} we analyze the effect of the inclusion of hyperons in the hadronic EoS.  For low values of $B$ the transition density is lower if the hadronic phase has hyperons, but for large values of $B$ transition density is larger if hyperons are present. This emphasizes the fact that the transition density depends not only on the stiffness of the equations of state, but has also a non-trivial dependence on the flavor composition of matter.

\section{Discussion and Conclusions}
\label{Discussion}

In this work we have analyzed the deconfinement transition from hadronic matter to quark matter and investigated the
role of color superconductivity, the effect of the trapped neutrinos at different temperatures, and the effect of hyperons in the hadronic phase.  To describe the strongly interacting matter a two-phase picture is adopted. For the hadronic phase we use different parametrizations of a non-linear Walecka model which includes hadrons, electrons,  and  electron  neutrinos  in  equilibrium  under weak interactions.  Just deconfined quark-matter is described as a gas composed by $u$, $d$ and $s$ quarks, electrons,  and electron neutrinos  which are \emph{not} in  equilibrium  under weak interactions. The equation of state of the quark phase is calculated within the frame of the MIT bag model including color superconductivity.
In order to determine the energy density at which deconfinement occurs we assume that the transition is of first order and
we impose flavor conservation during the transition in such a way that just deconfined quark matter is transitorily
out of equilibrium with respect to weak interactions. Additionally, the quark  phase must be color neutral.
Moreover, when color superconductivity is included together with flavor conservation,
the most likely configuration of the just deconfined phase is 2SC provided the pairing gap is large enough \citep{Lugones2005}. Thus, we assumed in this work that the just deconfined phase is 2SC and compared the results with the unpaired case and with results obtained within the NJL model.

Our results show that color superconductivity facilitates the transition for temperatures below $T_c$ (see Figs.~1 and 2).
This effect may be strong if the superconducting gap is large enough (Fig.~3).
Notice that the critical density $\rho^H$ for deconfinement tends to be small at high temperatures, it increases at intermediate $T$ and turns to decrease for  $T < T_c$ (see solid lines in Figs.~1 and 2).
As in previous work \citep{Lugones1998} we find that trapped neutrinos tend to increase the critical density $\rho^H$ for the formation of quark matter. However, if the just deconfined phase is in a color superconducting state, this effect is very small (see Fig.~4) in coincidence with the results obtained by \cite{LugonesCarmoNJL} within the NJL model. Concerning the presence of hyperons in the hadronic EoS we find that the transition density is lower (higher) for hadronic matter with hyperons if the quark matter EoS has small (large) values of $B$.  This behavior can not be understood solely in terms of the stiffness of the equations of state, but depends in a non-trivial way on the flavor composition of the hadronic and quark phases.

It is worth comparing the present results with those obtained in \cite{LugonesCarmoNJL} within the NJL model.
As explained in Sec.~2 it is possible to define within the NJL model a quantity $B'$ which plays a role similar to that of the bag constant $B$ in the MIT bag model. The \textit{set 1} and \textit{set 2} parametrizations of the NJL model employed in \cite{LugonesCarmoNJL} correspond
to $B$ = 353 MeV/fm$^{3}$  and 337.2 MeV/fm$^{3}$ respectively.
The behavior of the transition density $\rho^H$ as a function of $T$ is similar for both models. In the right panels of Figs.~1 and 2 of \cite{LugonesCarmoNJL}  we can also observe a ``belly'' in the curves that is analogous to the one observed in the solid lines of Figs.~1 and 2
of the present paper. With respect to neutrino trapping, in both models the transition density varies
within a few percent for  $\mu_{\nu_{e}}^{H}$ in the range 0--150 MeV. However, the qualitative behavior is somewhat different in both models
since there is a slight decrease of the transition density $\rho^H$ in the NJL case.
With respect to color superconductivity,
notice that in the case of the NJL model the pairing gap has a density dependent value that is determined by solving the gap equation
while in the here-used equation of state it has been considered as a free parameter. As expected, we find
in both cases that the transition density decreases at low temperatures because the pairing gap
increases (see Figs.~1 and 2 of the present paper and Figs.~1 and 2 of \cite{LugonesCarmoNJL}).
In Table~\ref{table2} we notice that there is an interesting coincidence in the numerical values of the transition density within the NJL and the MIT model.
The results are coincident within a $\sim 5 \%$, i.e. very similar in spite of the very different equations of state.

\begin{table}
\centering
\begin{tabular}{c|c|c}
\hline \hline
 T [MeV]        &     $\rho^{H} / \rho_{0}$         &   $\rho^{H} / \rho_{0}$      \\
                 & (GM1hyp$+\nu$ and MIT)  & (GM1hyp$+\nu$ and NJL)      \\
\hline
   2   & $ \;  7.6 \;$ (MIT $B337$)  &   $ \; 7.5 \;$ (NJL \textit{set 2}) \\
   2   & $ \; 10.0 \;$ (MIT $B353$)  &   $ \; 9.5  \;$ (NJL \textit{set 1}) \\
   30  & $ \;  8.7 \;$ (MIT $B337$)  & $ \;   8.2  \;$ (NJL \textit{set 2}) \\
   30  & $ \; 10.3 \;$ (MIT $B353$)  &  $ \; 10.0 \;$ (NJL \textit{set 1}) \\
\hline\hline
\end{tabular}
\caption{Comparison of the deconfinement transition density within the MIT bag model and the Nambu-Jona-Lasinio model.
Values for  the \textit{GM1hyp$+\nu$ and NJL} case have been extracted from Fig.~3 of  \cite{LugonesCarmoNJL}. Since the transition density depends very little on the chemical potential of trapped neutrinos we have adopted a typical value of $\rho^{H}$ in each case.  Note that the results agree within a few percent. }
\label{table2}
\end{table}

{The final configuration of the star after the transition depends on many issues that are out of the scope of the present paper, such as the presence or absence of mixed phases, the phase diagram of color superconducting quark matter in beta equilibrium, or the effect of rotation. In particular, we notice that the maximum allowed mass for each model is strongly affected by the distribution of the angular velocity in the radial direction and by the consequent degree of differential rotation \citep{Galeazzi}. In view of this, having $M_{max} > 2 M_{\odot}$ through the integration of a simple static Tolman-Oppenheimer-Volkoff equation should not be considered as a necessary condition for consistency of our calculations with observations. Anyway, we notice that some of the here used models give static stable stars when the quark phase is in the CFL state and no mixed phases are considered. For example, for $B$=80 MeV fm$^{-3}$ and large enough
$\Delta_0$ stable CFL strange stars with $M_{max} > 2 M_{\odot}$ are obtained by \cite{Lugones2003}. For stable static hybrid stars with $B$ = 337 and 353 MeV fm$^{-3}$ the reader is referred to \cite{Weissenborn2011} and references therein.}

A final comment is worthwhile concerning the formation of mixed hadron-quark phases, in which the electric charge is zero globally but not locally, i.e. the two phases have opposite charges \citep{gl92}. Mixed phases cannot form in the here studied just-deconfined phase \citep{Lugones1998}, because the flavor conservation condition guarantees that a just-deconfined quark-matter drop initially has exactly the same electric charge as the hadronic drop from which it originated (i.e. zero). Of course, charge separation could occur later on (if energetically preferred) and a mixed phase could form. However, notice that Debye screening effects and the surface tension can prevent mixed phases to form (see e.g. \cite{tatsumi,endo} and references therein). In any case, this study of mixed phases concerns the state of the system at times that are much longer than the ones that are addressed in this paper.

Within the MIT bag model, the expected effects on protoneutron star evolution are as follows.
When a PNS is formed it is hot and it has a large amount of trapped neutrinos. If color superconductivity were not considered, cooling will increase the transition density while deleptonization will decrease it \citep{Lugones1998}. Since both effects compete which each other it is possible that the transition is inhibited in the initial moments of the evolution of neutron stars \citep{Lugones1999}. As shown in the present paper, when color superconductivity is taken into account, the decrease of temperature decreases the transition density (due to the increase of the pairing gap).
Therefore,  both cooling and deleptonization of the PNS increase the probability of deconfinement as the PNS evolves.

\section*{Acknowledgments}
T. A. S. do Carmo acknowledges the financial support received from CAPES.
G. Lugones acknowledges the financial support received from FAPESP.


\end{document}